# Multi-Site Health Research Integrating Complementary Data Sources: A Scoping Review of Statistical Inference Methods for Vertically Partitioned Data

Marie-Pier Domingue[1,2,3], Simon Lévesque[1,2,3], Anita Burgun[3,4], Jean-François Ethier[1,5,6], Félix Camirand Lemyre[1,2,5]

**Author affiliations:**

[1]Groupe de recherche interdisciplinaire en informatique de la santé (GRIIS), Université de Sherbrooke, Sherbrooke, Canada

[2]Département de mathématiques, Université de Sherbrooke, Sherbrooke, Canada

[3]Institut Imagine, Université Paris Cité, Paris, France

[4]Department of Medical Informatics, Necker-Enfants malades Hospital, AP-HP, Paris, France

[5]Health Data Research Network Canada

[6]Département de médecine, Université de Sherbrooke, Sherbrooke, Canada

**Corresponding author:**

Jean-François Ethier

Groupe de recherche interdisciplinaire en informatique de la santé (GRIIS)
Faculté de médecine et des sciences de la santé
Université de Sherbrooke
3001, 12e Avenue Nord,
Sherbrooke (Québec) J1H 5N4
Canada
Phone: +1 (819) 821-8000, extension 74977
Email: jean-francois.ethier@usherbrooke.ca

**Abstract:**

*Background and objectives*

To address the multidimensional nature of health-related questions, advances in health research often require integrating information from various data sources within statistical analyses. When complementary information pertaining to the same set of individuals are distributed across different institutions, vertical methods make it possible to obtain analysis results without sharing or pooling individual-level data. To guide stakeholders toward a transparent and rigorous use of vertical methods with sensitive health data, this study aims to (1) Identify existing vertical methods enabling statistical inference (confidence interval estimation and hypothesis testing); and (2) Characterize the methodological properties of these methods and the current extent of their use with health data.

*Methods*

We conducted a scoping review following PRISMA-ScR using four interdisciplinary databases. We then systematically extracted the characteristics of identified vertical methods with respect to comparability with the pooled analysis, efficiency of communication schemes and confidentiality. We additionally screened studies that cited included articles to identify applications on vertically partitioned real-world health data.

*Results*

Among 2887 articles initially screened, 30 were included in the review, of which a majority mentioned health analytics. Inference for the linear and the logistic regression framework were the most frequent statistical inference tasks undertaken in proposed methods. Equivalence with the pooled analyses was not systematically addressed and most methods required multiple communications between participating parties. Almost all articles described their approach as privacy-preserving, although a minority provided privacy assessments. Very few published health studies were found to report the use these methods.

*Conclusion*

The scope of existing approaches enabling statistical inference for vertically partitioned data is still relatively limited. Most existing methods do not concurrently achieve results equivalent to centralized analyses, high communication efficiency, and guaranteed protection of individual-level data.

1. <u>INTRODUCTION</u>

Health sciences research increasingly requires performing analyses that integrate variables derived from clinical, genomic, socioeconomic, and environmental data sources; linking such complementary information pertaining to the same group of individuals is recognized as a key enabler for capturing the multidimensional nature of contemporary challenges [1,2]. For example, coupling electronic health records with genomic studies can lead to discoveries that will ultimately improve healthcare [3]. However, such data are generally distributed across multiple institutions, where strict legal, ethical and administrative regulations frequently prevent the transfer of individual-level data outside the organization that holds them. Regulations typically also restrict sharing of any information that could enable the reconstruction of individual-level data. As a concrete example, these challenges are currently faced by researchers in Quebec who wish to jointly analyze medico-administrative data held by the Institut de la statistique du Québec (ISQ) and health determinants, including nutritional behaviours and cognitive functions, held by the NuAge Database and Biobank [4].

Approaches in the field of distributed analytics have gained momentum to safely enable analysis of sensitive health data from multiple sources [5]. These approaches rely on the coordinated exchange of intermediate numerical outputs across participating organizations instead of pooling individual-level data for performing data analyses. Conceptually, two types of data *partitions* are typically distinguished in the literature for a dataset: the *horizontal partition*, where organizations—called *nodes* hereafter—possess the same variables for distinct individuals; and the *vertical partition,* where complementary subsets of variables for the same set of individuals are held by different nodes (see Figure 1).

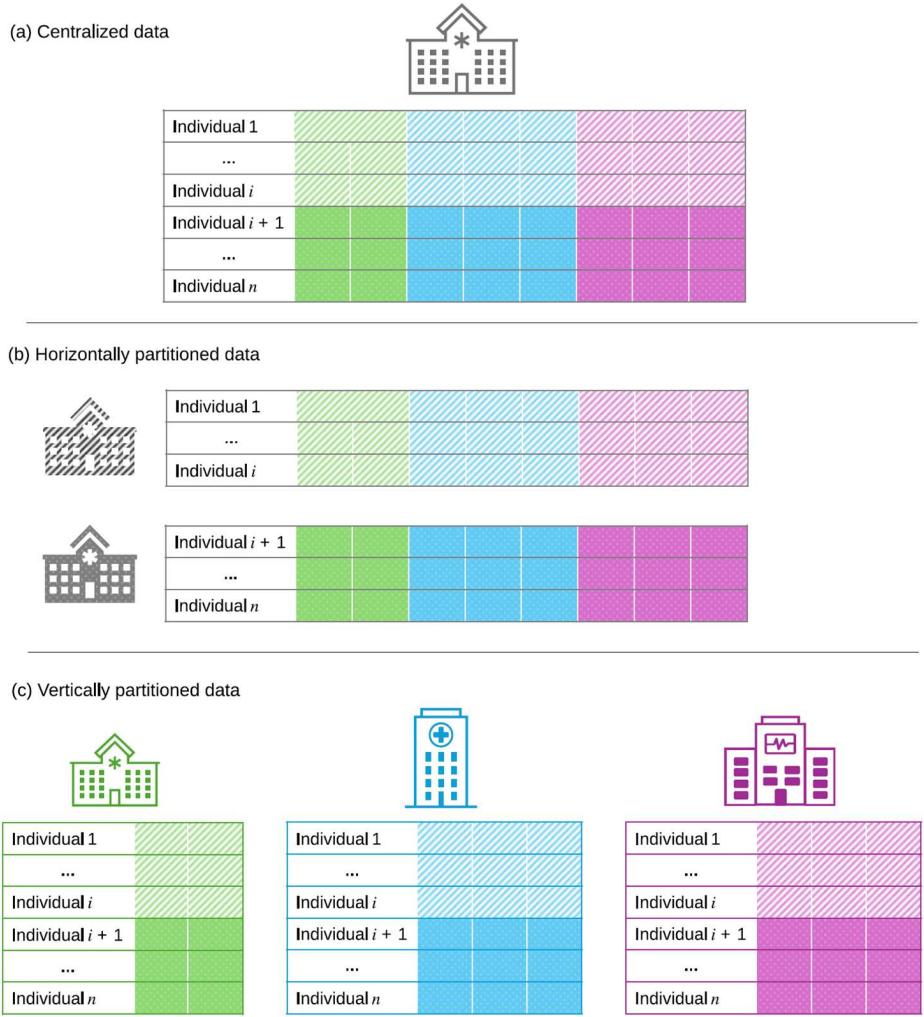

*Figure 1. (a) Centralized data. (b) Horizontally partitioned data (example of two data nodes), where each pattern represents a different data node. (c) Vertically partitioned data (example of three data nodes), where each color represents a different data node.*

In recent years, the horizontal distribution setting has received significant attention, largely driven by the wide adoption of electronic health records and the emergence of multi-machine strategies to accommodate massive data [6–9]. In comparison, the vertical distribution setting has received considerably less attention despite its high potential to address multidimensional and multimodal analyses [10–13]. An explanation lies in the fact that the vertical setting has often been reported as inherently more challenging [14–16]; while in the horizonal case every node possesses all variables required to adjust local models, such a property does not hold with vertically partitioned datasets (see Figure 1).

Previous review work in the vertical distribution setting has been centered around machine learning [17,18]. More general reviews for distributed data regardless of their partition type have also been conducted [16,19], including some specifically focused on partitioned health data [5,10,20]. While obviously important, these reviews did not focus on statistical inference tasks [21,22]—defined in this work as performing confidence intervals estimations and/or hypothesis testing—which continue to play a significant role in present-day health research practices, e.g. to assess treatment effect or explore associations between a plurality of factors and health outcomes.

At this stage, the extent of the availability of vertical statistical analytics methods has not yet been determined and the degree to which these existing methods can be implemented with real-world health databases is unclear. This gap warrants careful consideration to support health analytics and guide decisions regarding collaborative research.

Given the ubiquity of statistical inference tasks in health research, this paper seeks to delineate the scope of existing methods that pertain to vertical statistical inference in view of their potential applicability to health data analysis to inform their utility for health data analysts and providers. More specifically, the first objective is to identify existing vertical methods enabling statistical inference. To guide the critical use of these methods in future applications, the second objective is to provide a systematic account of the methodological properties of existing methods, with respect to features highlighted as supporting their suitability for real-world analyses (detailed below), and to characterize the current extent of their use with health data.

To achieve the first objective, we employed a scoping review methodology. For the second objective, we focused on method-specific features often deemed central to decision-making for health analytics stakeholders [9,23] (see Figure 2) and that are known to differ markedly across distributed methods [17], namely:

1. Comparability with the pooled analysis results, defined as the precision of results produced by a distributed method relative to those obtained from the analysis of the same data in a context where those data would have been centralized [9,24], with commonly targeted levels of comparability including exact recovery (the gold standard) and equivalence of results up to numerical tolerance;
2. Efficiency of communications schemes, to consider the burden of large communication volumes that has been reported as a concern for vertical analyses [15,25], especially when manual revision of exchanged quantities might be required;
3. Confidentiality of individual-level data, as theoretical guarantees that individual-level data cannot be retrieved are critical for many organizations responsible of protecting the privacy of data they hold [5], and the distributed nature of a method does not inherently imply such guarantees [26,27].

To assess the extent of the use of identified methods, we additionally explored whether studies based on real-world health data reported the use of these methods.

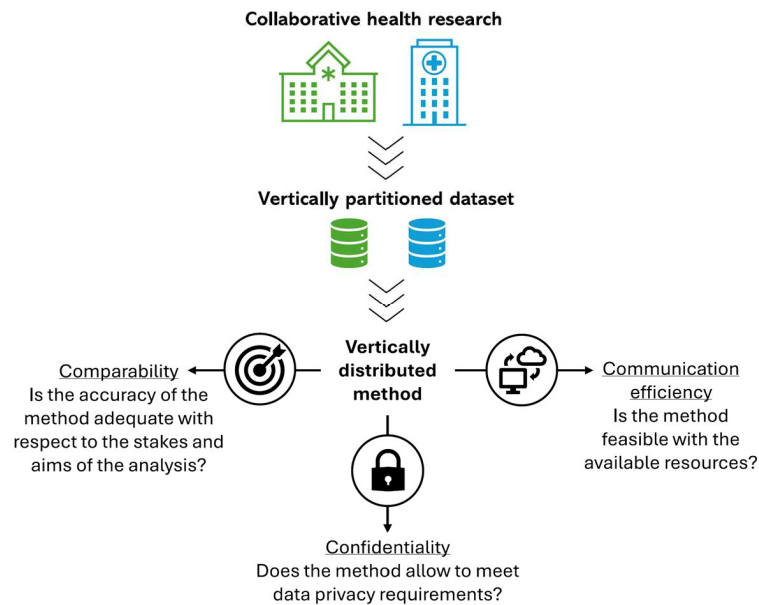

*Figure 2. General process in collaborative health data research with vertically partitioned data and main features fluctuating across methods used to guide the classification process.*

2. METHODS

**2.1. Methodology related to objective 1**

Scoping reviews allow to derive clear indications of the volume of studies available on a specific topic while reporting key characteristics related to the topic, which was well suited for our objectives [28]. To ensure a systematic and reproducible process, the method prescribed by Arksey and O'Malley [29] was adopted, and the PRISMA-ScR (PRISMA extension for scoping reviews) guidelines were followed [30]. The complete scoping review protocol is available in Appendix A.

### 2.1.1. Research question

In accordance with the Arksey and O'Malley framework, the research question was explicitly formulated prior to the literature search as follows: "What existing distributed analytics methods allow conducting statistical inference procedures with vertically partitioned data?". The following sub-question was added to comprehensively address our objectives: "What are the characteristics specific to each of these methods to provide systematic categorization for health data providers?".

### 2.1.2. Inclusion and exclusion criteria

To be included, an article had to present a solution to perform inferential statistics with vertically partitioned data. Studies that presented a method that applies to horizontally partitioned data only, or studies for which the method required pooling all individual-level data, were therefore not included. We excluded papers that did not provide new methodological contributions. Any article not published or not available in English or French was excluded.

### 2.1.3. Keywords, search strategy and selection process

Based on the research question, we specified two components to guide keywords definition and elaborate research queries: (1) vertically partitioned data and (2) statistical inference. We drew from a previous scoping review from our research group pertaining to horizontally partitioned data [6] to establish keywords that would be appropriate for the inferential statistics component. More specifically, we conducted snowballing keywords search [31] among included articles from this previous review and validated that these keywords could be transferred to the vertically partitioned setting using papers already within our knowledge. Identifying keywords related to vertically partitioned data was more complex, as *vertical* was not systematically used in the literature. To ensure comprehensive screening, we defined broad keywords associated with this setting including *Distributed,*

*Partitioned* and *Privacy-preserving*, which inevitably captured articles beyond the vertical setting.

To address the interdisciplinary extent of the research question covering health analytics and statistics/computer sciences, the following abstract and citation databases were selected [32]: (1) Medline, (2) Scopus, (3) IEEE Xplore, and (4) ACM Digital Library. As we only aimed at published articles, we did not conduct any search among grey literature. Results were imported into the Covidence systematic review tool (http://www.covidence.org) for duplicate removal, paper selection and classification of selected articles.

Title and abstracts were screened by a single reviewer using a conservative approach; if the abstract could not lead to an irrefutable exclusion decision, the article was selected for full-text review. All selected articles were assessed through full-text screening by two independent and blinded reviewers. Reference screening of all included articles was additionally conducted to ensure the capture of articles that may have used different keywords to address this interdisciplinary topic.

### 2.2. Methodology related to objective 2

A classification template was elaborated (see Appendix A for details) to systematically extract information in alignment with Figure 2. As generic information, whether the method could be applied to block-partitioned data combining horizontal partitioning across individual subsets and vertical partitioning across feature holders was reported. Any mention of health analytics in the article was also noted.

To address comparability with the pooled analysis results, the type of model and/or inference task that could be conducted using the proposed method was extracted, along with the general methodological approach for the vertical setting, i.e. the core concepts allowing vertically partitioned computations. To provide an understanding of communication efficiency, the number of communications between nodes and whether an external third party was required were extracted.

The following characteristics were extracted to provide a complete picture of how confidentiality of individual-level data was covered:

- Whether privacy was at all mentioned in the article;
- Whether the method required sharing raw variables;
- Whether the concept of retrieving individual-level data was discussed;
- Whether any type of privacy assessment was provided. This is distinct from simply not directly sharing line-level data. It has been demonstrated that some methods may not prevent retro-engineering of line-level data by third parties [26].

Additional information that helped provide a more complete understanding of the method for health data stakeholders were extracted, including comparisons between the proposed method and the pooled approach. Data extraction was conducted by two independent reviewers.

Finally, studies that cited an included article were screened, and those that provided an application based on real-world health data were reported.

3. RESULTS

A total of 30 articles were included in the scoping review among the 2887 initially screened by the research queries, as presented in Figure 3. After removing 324 articles identified as duplicates, 2105 articles were identified as irrelevant according to title and abstracts screening. Following the full-text review, 25 articles were selected. Reference screening identified 5 additional articles that offered methodological contribution relevant to statistical inference with vertically partitioned data, although these articles did not explicitly use our predefined statistical inference keywords or did not explicitly mention distributed analyses. The list of all included articles is available in Appendix B.

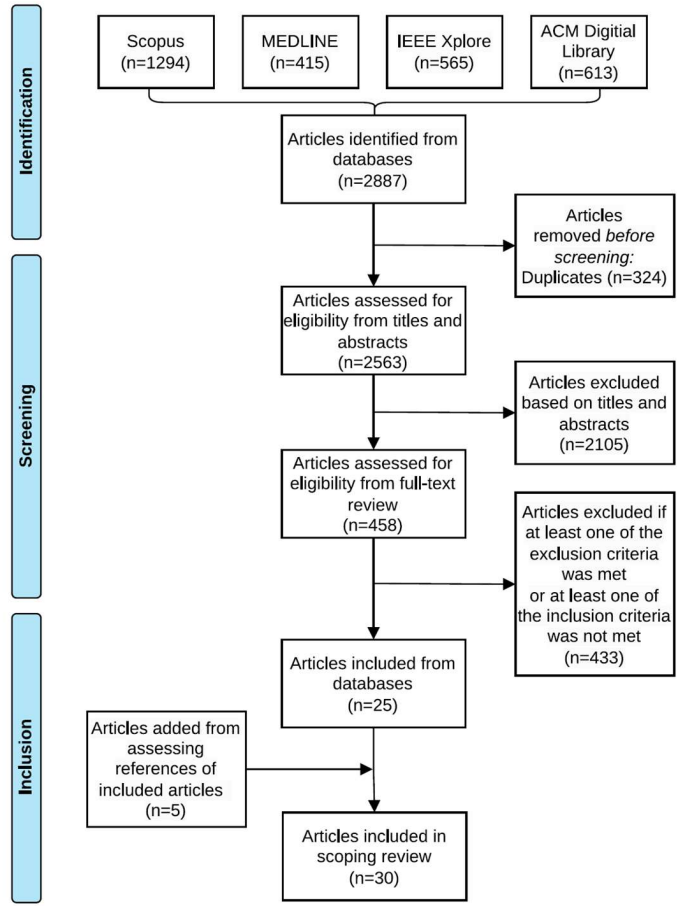

*Figure 3. PRISMA flow chart for article screening process. Detailed inclusion and exclusion criteria are described in the protocol.*

### 3.1. Description of included articles

Included articles were published after 2000, and a majority (n=21, 70.0%) was published after 2010. A subset of articles (n=9, 30.0%) addressed mixed partitions where features and individuals can be simultaneously partitioned, whereas the remainder focused solely on vertically partitioned data. Six of the included articles (n=6, 20.0%) [33–38] did not mention health analytics nor present any examples pertaining to health analytics.

### 3.2. Method characteristics pertaining to comparability with pooled analyses

#### 3.2.1. Types of models and inference tasks

In Table 1, we indicate whether the inferential tasks undertaken by each proposed method were specific to a regression setting and identify the statistical models addressed, noting that some papers covered more than one model or inferential task.

Table 1. Type of inference tasks addressed among included articles

| Type of inference tasks addressed | Articles[1] | Example pertaining to health analytics |
| --- | --- | --- |
| Inference under the linear regression model | [36–44] | *Identify significant factors, including duration of stay and age, associated with hospital expense as outcome* [41] |
| Inference under the logistic regression model | [33,34,45–48] | *Identify significant health factors associated with death as binary outcome* [46] |
| Inference under the Cox regression model | [14,25,49] | *Investigate the impact of health factors on survival probabilities among leukemia patients (survival analysis)* [14] |
| Inference under other types of parametric / semiparametric regression models | [38,50–52] | *Analyse the effect of smoking on metabolites that are upstream determinants of cardiovascular health* [51] |
| Other types of inference tasks (non-parametric two-sign test, Welch t-test, chi-square test, and others) | [35,53–60] | *Examine the impact of exposure to a risk factor (e.g. smoking) over the risk of a disease (e.g. cancer)* [55] |

[1]Total does not necessarily equal 30, because an article might present methods that apply to more than one model or inference task.

Among included articles, the linear regression framework (n=9, 30.0%) and the logistic regression framework (n=6, 20.0%) were the most frequently addressed. Three articles addressed the Cox regression model (n = 3, 10.0%), and four addressed other types of regression (n = 4, 13.3%). Some methods did not explicitly provide steps for the computation of standard errors and/or confidence intervals as part of the procedure (see e.g. [45]) but these quantities could be directly derived from computed quantities. Among the articles classified under *Other types of inference task* (n=9, 30.0%), structural equation models [54], Gaussian likelihood–based inference (encompassing, but not limited to, linear models and structural equation models) [53], nonparametric sign tests [35,57], causal inference [56,60], chi-square tests [55,58] and Welch t-tests [59] were addressed.

One method classified as addressing logistic regression focused specifically on Exact (finite-sampled) logistic regression [47], an approach that operates without relying on asymptotic approximations and applied in that study to a covariate-aware genome-wide association analysis.

*3.2.2. Methodologies for distributed procedures*

Three categories of techniques were used to address the vertical partitioning setting: encryption-based methods, methods based on vertically separable quantities and methods based on secure multiparty computations (SMC). We refer to Appendix B for a technical description of these approaches and Table B.2 for a classification of the primary technique used in included articles. The most frequently used primary technique, defined as the technique around which the method is articulated, was encryption-based protocols (n=13, 43.3%), followed by SMC (n=11, 36.7%) and vertically separable quantities (n=4, 13.3%).

*3.2.3. Comparison to the centralized analyses*

Comparability with the centralized analysis was sometimes reported, though the basis for comparison varied across papers, as reported in Table 2. Numerical comparisons between the distributed method and the true coefficients of the data-generating model were sometimes conducted using simulation studies (e.g. see [33,48]).

Table 2. Basis for comparison to the centralized setting

| Terminology employed | Article | Citation example |
|---|---|---|
| *Numerical agreement* | [44,48] | "The regression coefficients computed with our secure protocol agree to the 5th decimal digit with regression coefficients computed without any security." [44] |
| *Equivalency* | [38,49,53] | "It shows that VERTICOX achieves equivalent results for statistical tests in comparison to the centralized realization. VERTICOX does not change *p*-values and the confidence intervals." [49] |
| *Competitivity* | [25] | "Regarding the accuracy, the proposed DC-COX shows higher performance than the local party |

|  |  | analysis and competitive to a federated learning-based method and the centralized analysis." [25] |
| --- | --- | --- |
| *Identical/Same result or conclusion* | [14,40,41,53,54] | "We show that under a certain set of assumptions our method for estimation across these partitions achieves identical results as estimation with the full data."[53] |

Regarding the conclusions drawn from the basis of comparison, it was noted in [14] that a method yielding the same statistical conclusion as the centralized approach does not necessarily produce numerically identical estimates. Also, in [41,53], theoretical equivalence results were stated but numerical errors entailed by sequential computations were reported in numerical experiments. In addition, although [40] reported obtaining identical coefficients, the numerical results provided did not appear to fully support this claim.

Although certain methods relied on distributed algorithms designed to replicate the computations of the centralized procedure, comparability with the centralized setting was not always explicitly discussed [34,36,37,42,43,45,58]. Encryption protocols often relied on approximations affecting analytics results, and approximation errors were sometimes investigated, see e.g. [14,39,59]. In [56], data transformations and additive noise were used and were shown to impact accuracy. In [48], the original formulation of the optimization problem to solve was modified to create vertically separable quantities, although theoretical arguments supporting the equivalence between the results produced and the centralized procedure were not provided.

### 3.3. Method characteristics pertaining to efficiency of communication schemes

#### 3.3.1. Description of communication schemes

Six articles (n=6, 20.0%) proposed methods that were classified in this review as one-shot procedures [47,51,52,56,58,60], defined here as procedures using at most one outgoing and one incoming communication per data node, except for the data node fulfilling a role similar to that of a coordinating center (if applicable). Some methods attained the one-shot property only in the setting with two data nodes, see for example [41].

Among included articles, most methods relied on extensive iterative communication schemes, sometimes attributable to parameter updates during the optimization procedures (see e.g. [45]) and sometimes to encryption protocols (see e.g. [36]). Many methods based on SMC and encryption methodologies required a series of node-to-node communications among all participating data nodes (e.g. see [45,48]).

For methods pertaining to linear or logistic regression—two of the most frequently used approaches in health—we further examined the communication workflows required for statistical inference. For linear regression, identified methods used at least one back-and-forth communication per pair of data nodes (Figure 4b), and while no method could be broadly classified as one-shot, the one-shot property was attained under the specific setting of two data nodes in [41,43] (Figure 4a). For logistic regression, the method addressing Exact logistic regression was the only one identified as one-shot [47], although it should be noted that this method applies only to analyses involving two data nodes. The proposed procedures for the standard logistic regression model implied parameter updates through an iterative scheme, with most methods relying on a communication workflow similar to that shown in Figure 4c. More complex communication workflows were also encountered for the logistic regression model, including the procedure in [48] using a coordinating center as illustrated in Figure 4d.

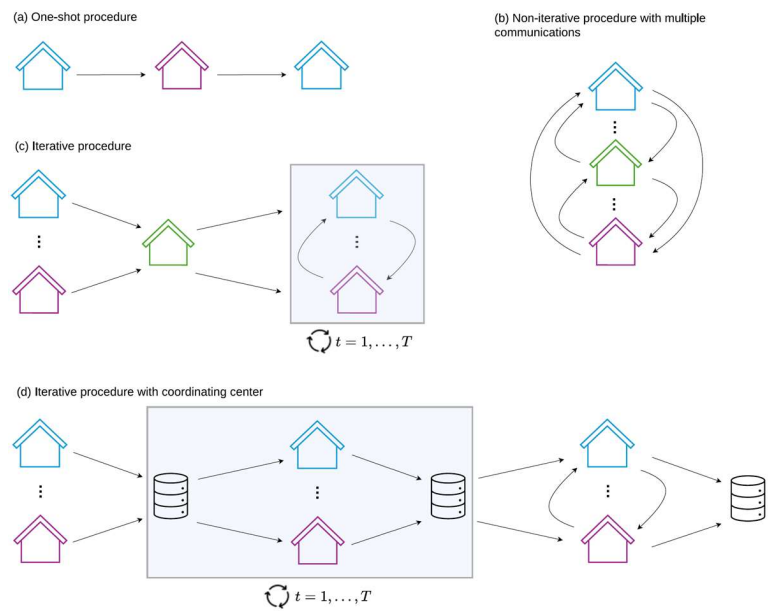

Figure 4. Main communication workflows for the vertically distributed linear regression and logistic regression models.

*3.3.2. Data nodes, third party and/or coordinating center*

Methods from eight included articles (n=8, 26.7%) required an external coordinator, sometimes referred to as master [25], server [48,49,54,58], aggregator [60], initializer [44] or central node [53]. While most methods considered the general case with an arbitrary number of participating data nodes, some focused specifically on the setting with two data nodes only [35,37,46,47,55,57–59].

**3.4. Method characteristics pertaining to confidentiality of individual-level data**

Two included articles [50,52] did not explicitly discuss privacy considerations, which is consistent with the framework adopted in these papers, where vertical partitioning was introduced as a computational strategy for handling massive datasets under a divide-and-conquer principle, rather than as a response to data-sharing constraints.

*3.4.1. Methods requiring the sharing of individual-level information*

Five articles (n=5, 16.7%) proposed methods relying on the assumption that at least one variable was made available to all data nodes. For instance, the response variable was assumed to be made available to multiple data nodes in linear [37], logistic [48] and Cox regression [25,49] settings. The method in [56] assumed that the treatment exposure, and potentially additional covariates, were available to all participating data nodes.

*3.4.2. Retrieving individual-level data and privacy assessments*

Except [50,52], all included articles used words such as *Secure, Privacy-preserving, Private* to qualify their proposed method and/or algorithms. The role of privacy among included articles is detailed in Figure 5.

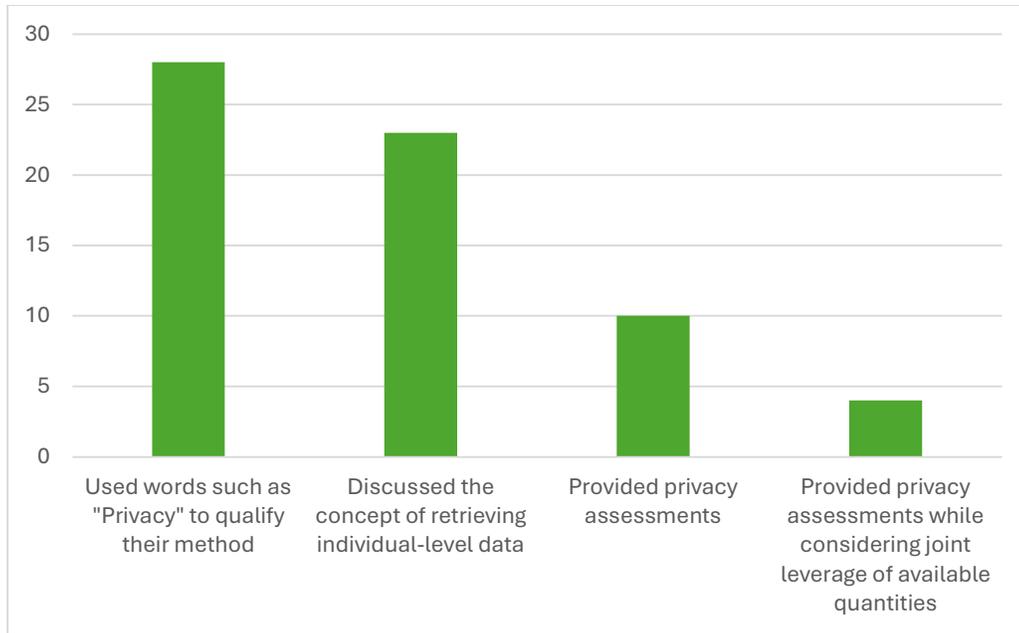

*Figure 5. Role of privacy among included articles.*

Some articles did not discuss the potential for reverse-engineering or recovering raw data from intermediate numerical outputs, and focused solely on the fact that individual-level data were not directly exchanged [36,37,49,51,56]. Many other articles went beyond noting that individual-level data were not exchanged and discussed the possibility of reconstructing such data from intermediate numerical outputs but without providing any extensive privacy assessment or offering any theoretical guarantees that individual-level data could not be retrieved [25,33,34,38,39,41–43,45,46,48,53,54].

Elaborating further on the concept of retrieving individual-level data, some articles provided privacy assessments based on encryption theory [14,35,40,44,47,55,57–60]. Among these, some authors characterized each individual operation as private [14,40,55,59], in the sense that the intermediate numerical outputs produced by any single operation were shown not to compromise individual-level data confidentiality. Whether the sequence of intermediate outputs disclosed could be jointly exploited to infer additional information regarding individual-level data was addressed in a few articles only [35,44,47,57]. Among all articles providing privacy assessments, the prevailing assumption was the *honest-but-curious*, also referred to as *semi-honest*, where participating nodes are assumed to execute the protocol as prescribed without colluding with other entities, but are still curious to infer individual-level data using the quantities they hold during and after the procedure (see e.g., [57,61] for details).

### 3.5. Applications of vertical analytics methods

We reported works that cited included articles and explicitly presented real-world cases of health applications or platforms intended for distributed health analytics with vertically partitioned data. First, an article discussing PopMedNet [62], a software platform for automated distributed analytics, explicitly cited [34,43], as it implemented versions of the vertical linear and logistic regression methods proposed therein. The same article also cited [49], identifying its Cox regression method as a potential future extension. This software was reported to be used by several large, distributed health data networks. Another system named D-CLEF [63], developed to enhance collaborative health research, noted to have incorporated the vertical method for logistic regression proposed in [48]. Finally, the method for linear regression proposed in [41] was used to study stress using heart rate and sleep data from medical devices as well as overtime hours from labor association [64].

## 4. DISCUSSION

In this review, we reported existing methods to conduct statistical inference tasks with vertically partitioned data, provided a systematic account of their methodological properties and assessed the extent of their use in with real health data.

Among the 30 included articles, the proposed methods mainly covered the linear regression model, the logistic regression model and the Cox regression model, as well as other statistical analyses such as Welch and chi-square tests. While these unlock potential analyses with health data, a gap remains for a variety of semi-parametric and non-parametric models, and even for a large class of parametric generalized linear models. Some models widely used in health analytics remain unaddressed, including generalized linear mixed models, which are commonly adopted in clinical studies to account for correlated observations [65]. These models may be particularly relevant to analyze longitudinal data from ISQ and NuAge.

Regarding comparability with centralized analyses, theoretical and numerical equivalence were not systematically addressed, and the basis for comparison varied across papers. Most methods aimed to achieve results identical to those of the centralized procedure by replicating its algorithmic iterations in a distributed setting. Equivalence was sometimes deemed to have been achieved based on numerical experiments showing agreement with the centralized procedure up to a specified decimal precision.

Regarding communication schemes efficiency, many methods relied on heavy communications workflows due to SMC protocols, which may limit feasibility for institutions that do not possess automated infrastructures, or contexts that require manual revision of exchanged quantities. When more than two participating data nodes were involved, hardly any methods offered a one-shot communication scheme. Health institutions often have limited time and resources to conduct analyses while still aiming for the gold standard in terms of result accuracy, which rules out most existing methods for the logistic and Cox regression models, as they entail iterative communication schemes. Methods addressing models admitting a closed-form solution, such as the linear regression model, often yielded results equivalent to those of centralized analyses while remaining communication-efficient. We note that the computational complexity entailed by encryption protocols was highlighted as a barrier for real-world applications by multiple authors (see e.g., [14]).

While privacy appeared to be the main motivation of most included articles, some methods relied on the assumption that one or multiple variables were available or shared across all data nodes, which represents a major limitation for their use in health analytics. Also, in some cases, methods were described as privacy-preserving or secure on the basis that their underlying procedure did not require sharing individual-level data, without additional assessment of privacy guarantees. However, for many organizations dealing with sensitive individual-level data, such as Statistics Canada and ISQ, the absence of direct sharing of individual-level data does not, in itself, constitute a guarantee of privacy protection. The absence of formal privacy guarantees beyond the non-sharing of individual-level data should be carefully weighed by data providers and other stakeholders before deploying such methods in practice, particularly when sensitive health data are involved.

These assessments should further explore the risk of disclosing final estimates or associated statistical results, as this was rarely addressed among included articles. In some cases, it was explicitly assumed that the results (estimates, confidence intervals) did not constitute sensitive information that could be leveraged along previously exchanged quantities.

Although most reported articles mentioned health analytics and many reported proofs-of-concept on artificially distributed health data, very few papers appeared to report real-world applications of these methods. While the present scoping review does not permit definitive conclusions regarding such scarcity, this limited uptake may point to unresolved practical challenges. In our experience collaborating with data custodians, several practical important considerations often arise, including the lack of confidentiality guarantees, the limited range of statistical models supported by existing methods, which may preclude performing analyses required in practice, and the operational burden associated with communication-intensive procedures. This review suggests that these aspects remain only partially addressed in the current literature. Although further investigations involving data custodians would be necessary to comprehensively understand barriers to applicability, addressing these considerations could represent an important step towards broader adoption of vertical methods in collaborative health research.

### 4.1. Limitations

This review did not address general operational challenges entailed when conducting vertically partitioned analyses, such as the requirement for individual-level data alignment. These types of macro-level considerations need to be addressed before implementing a given method, but do not constitute method-specific features in a statistical sense. Among the challenges faced during the research process, the use of broad keywords for distributed partitions led to a high number of articles to screen—a lot of which only pertained to horizontally partitioned data and had to be excluded—but was inevitable to ensure a comprehensive search due to the wide and varying vocabulary associated with this topic.

### 5. CONCLUSION

This study highlights that the scope of existing approaches enabling statistical inference for vertically partitioned data is still relatively limited, and the majority of existing methods cover statistical inference tasks for standard parametric models. Most existing methods do not concurrently achieve results equivalent to centralized analyses, high communication efficiency, and guaranteed protection of individual-level data, which limits their applicability in health research. Protection of individual-level data confidentiality was often only partially addressed, and additional, in-depth, privacy assessments may be necessary to enable the application of these methods in settings involving sensitive health data and data providers requiring formal privacy guarantees.

*CRediT authorship contribution statement*

**Marie-Pier Domingue:** Conceptualization, Methodology, Investigation, Writing – original draft, Writing – review and editing, Visualization; **Simon Lévesque**: Methodology, Investigation, Writing – original draft, Writing – review and editing; **Anita Burgun:** Conceptualization, Writing – review and editing, Supervision; **Jean-François Ethier**: Conceptualization, Methodology, Writing – original draft, Writing – review and editing, Supervision; **Félix Camirand Lemyre**: Conceptualization, Methodology, Writing – original draft, Writing – review and editing, Supervision

*Funding sources*

This work was supported by the Natural Sciences and Engineering Research Council of Canada – Discovery Grant; the Health Data Research Network Canada, an initiative funded by the Canadian Institutes of Health Research; the Chaire de recherche en informatique de la santé de l'Université de Sherbrooke, and the Chaire de recherche en méthodologies statistiques appliquées aux enjeux de santé et phénomènes sociaux de l'Université de Sherbrooke. Marie-Pier Domingue received a scholarship from the Natural Sciences and Engineering Research Council of Canada. Simon Lévesque received a scholarship from the Natural Sciences and Engineering Research Council of Canada. Jean-François Ethier received a Clinical Research Scholar Junior 2 from the Fonds de recherche du Québec – Santé. This work was supported by State funding from the Agence Nationale de la Recherche under "Investissements d'avenir" program (ANR-10-IAHU-01).

*Declaration of competing interest*

The authors declare that they have no known competing financial interests or personal relationships that could have appeared to influence the work reported in this paper.

*Appendix A. Scoping review protocol and extraction template*

*Appendix B. Supplementary Tables*

*Figure captions*

Figure 1. (a) Centralized data. (b) Horizontally partitioned data (example of two data nodes), where each pattern represents a different data node. (c) Vertically partitioned data (example of three data nodes), where each color represents a different data node.

Figure 2. General process in collaborative health data research with vertically partitioned data and main features fluctuating across methods used to guide the classification process

Figure 3. PRISMA flow chart for article screening process. Detailed inclusion and exclusion criteria are described in the protocol.

Figure 4. Main communication workflows for the vertically distributed linear regression and logistic regression models

Figure 5. Role of privacy among included articles

*Multi-Site Health Research Integrating Complementary Data Sources: A Scoping Review of Statistical Inference Methods for Vertically Partitioned Data - APPENDIX A.* **Scoping Review Protocol**

**TABLE OF CONTENT**

1. **RESEARCH QUESTION**

    1.1. **What existing distributed methods allow conducting statistical inference procedures with vertically partitioned data?**
    - *Regarding: Methods for different statistical models; Methods for various settings in terms of privacy; Methods for different data settings (e.g. continuous vs binary covariates).*

    1.2. **What are the characteristics specific to each of these methods to provide systematic categorization for health data providers?**
    - *Regarding: Quantities exchanged; Communication workflow; Settings for nodes and central server; Capacity to reach exact estimates from centralized analysis; Types of variables; Targeted application setting (if any); Privacy analysis.*

2. **METHODS**

## 2.1. Keywords

In accordance with our previous review, while making sure we capture all potential existing methods, two categories of keywords were targeted, corresponding to the two themes in the research question.

- Distributed analyses:
    - Partitioned
        - Partitioned
        - Federated
        - Distributed
        - Aggregated
        - Privacy-preserving
        - Multiparty
        - Multiple sources
    - Data/Analyses
        - Dataset
        - Database
        - Inference
        - Estimation
        - Regression
        - Statistical Analysis
        - Survival Analysis
        - Analysis
        - maximum likelihood
- Statistical inference:
    - statistical inference
    - statistical analysis (analyses)

- statistical estimation
- confidence interval
- hypothesis test, hypothesis testing
- significant coefficient
- significant parameter
- standard error

### 2.1.1. Limits and restrictions

While the research question was focused on vertically partitioned data, existing work seemed to be inconsistent regarding the choice of keywords to denote such partitioning. The broader topic of *distributed analyses* was therefore targeted, even though this implied more noise in the results. This can be seen as a methodological strategy to increase the sensitivity, which implied lower specificity.

## 2.2. Research strategies

The following abstract and citation databases have been selected: (1) Medline, (2) Scopus, (3) IEEE Xplore, and (4) ACM Digital Library. This choice of databases was motivated by the interdisciplinary extent (fields of statistics and health) of the research question.

The previously mentioned keywords were combined to define the following research strategies among all four databases, also in collaboration with a specialist in documentary research.

### 2.2.1. Medline

*((AB ((("privacy-preserving" OR "federated" OR "Distributed" OR "aggregated" OR "partitioned" OR "multiple sources" OR "multiparty") N2 ("Dataset\*" OR "Database\*" OR "inference" OR "Estimat\*" OR "Statistical Analy\*" OR "Regression\*" OR "Survival Analysis" OR "maximum likelihood" OR "vertically")) NOT ("distributed lag" OR "normally distributed" OR "signal\*" OR "sensor\*" OR "electric\*")) OR SU ((("privacy-preserving" OR "federated" OR "Distributed" OR "aggregated" OR "partitioned" OR "multiple sources" OR "multiparty") N2 ("Dataset\*" OR "Database\*" OR "inference" OR "Estimat\*" OR "Statistical Analy\*" OR "Regression\*" OR "Survival Analysis" OR "maximum likelihood" OR "vertically")) NOT ("distributed lag" OR "normally distributed" OR "signal\*" OR "sensor\*" OR "electric\*")) OR TI ((("privacy-preserving" OR "federated" OR "Distributed" OR "aggregated" OR "partitioned" OR "multiple sources" OR "multiparty") N2 ("Dataset\*" OR "Database\*" OR "inference" OR "Estimat\*" OR "Statistical Analy\*" OR "Regression\*" OR "Survival Analysis" OR "maximum likelihood" OR "vertically")) NOT ("distributed lag" OR "normally distributed" OR "signal\*" OR "sensor\*" OR "electric\*"))) AND TX("statistical inference" OR "confidence interval\*" OR "Statistical Estimat\*" OR "hypothesis test\*" OR "significant coefficient\*" OR "significant parameter\*" OR "standard error\*" OR "Statistical Analy\*" OR "p-value\*" OR "intermediary statistics" OR "intermediate statistics"))*

### 2.2.2. Scopus

*TITLE-ABS-KEY (((((("privacy-preserving" OR "federated" OR "Distributed" OR "aggregated" OR "partitioned" OR "multiple sources" OR "multiparty") W/2 ("Dataset\*" OR "Database\*" OR "*

inference" OR "Estimat*" OR "Regression*" OR "Statistical Analysis" OR "Survival Analysis" OR "Analysis" OR "maximum likelihood" OR "vertically"))) AND ("statistical inference" OR "confidence interval*" OR "Statistical Estimat*" OR "hypothesis test*" OR "significant coefficient*" OR "significant parameter*" OR "standard error*" OR "Statistical Analy*" OR "p-value*" OR "intermediary statistics" OR "intermediate statistics")) AND NOT ("distributed lag" OR "normally distributed" OR "sensor*" OR "electric*" OR "signal*"))

### 2.2.3. IEEE Xplore

(((((("Document Title":"privacy-preserving" OR "Document Title":"federated" OR "Document Title":"Distributed" OR "Document Title":"aggregated" OR "Document Title":"partitioned" OR "Document Title":"multiple sources" OR "Document Title":"multiparty") NEAR/2 ("Document Title":"Dataset*" OR "Document Title":"Databases" OR "Document Title":"Database" OR "Document Title":"inference" OR "Document Title":"Estimat*" OR "Document Title":"Regression" OR "Document Title":"statistical Analysis" OR "Document Title":"survival Analysis" OR "Document Title":"maximum likelihood" OR "Document Title":vertically)) OR (("Abstract":"privacy-preserving" OR "Abstract":"federated" OR "Abstract":"Distributed" OR "Abstract":"aggregated" OR "Abstract":"partitioned" OR "Abstract":"multiple sources" OR "Abstract":"multiparty") NEAR/2 ("Abstract":"Dataset*" OR "Abstract":"Databases" OR "Abstract":"Database" OR "Abstract":"inference" OR "Abstract":"Estimat*" OR "Abstract":"Regression" OR "Abstract":"statistical Analysis" OR "Abstract":"survival Analysis" OR "Abstract":"maximum likelihood" OR "Abstract":vertically))) AND ("Full Text & Metadata":"statistical inference" OR "Full Text & Metadata":"confidence interval*" OR "Full Text & Metadata":"Statistical Estimat*" OR "Full Text & Metadata":"hypothesis test*" OR "Full Text & Metadata":"significant coefficient" OR "Full Text & Metadata":"significant parameter" OR "Full Text & Metadata":"standard error*" OR "Full Text & Metadata":"Statistical Analysis" OR "Full Text & Metadata":"p-value" OR "Full Text & Metadata":"intermediary statistics" OR "Full Text & Metadata":"intermediate statistics")) NOT ("Abstract":"distributed lag" OR "Abstract":"sensors" OR "Abstract":"electrical" OR "Abstract":"signal*"))

### 2.2.4. ACM Digital Library

(((((Abstract: "privacy-preserving" OR Abstract: "federated" OR Abstract: "distributed" OR Abstract: "aggregated" OR Abstract: "partitioned" OR Abstract: "multiple sources" OR Abstract: "multiparty") AND (Abstract: dataset* OR Abstract: database* OR Abstract: "inference" OR Abstract: "estimation" OR Abstract: "statistical analysis" OR Abstract: "regression" OR Abstract: "survival analysis" OR Abstract: "maximum likelihood")) OR Abstract: "vertically partitioned") OR (((Title: "privacy-preserving" OR Title: "federated" OR Title: "distributed" OR Title: "aggregated" OR Title: "partitioned" OR Title: "multiple sources" OR Title: "multiparty") AND (Title: dataset* OR Title: database* OR Title: "inference" OR Title: "estimation" OR Title: "statistical analysis" OR Title: "regression" OR Title: "survival analysis" OR Title: "maximum likelihood")) OR Title: "vertically partitioned")) AND ("statistical inference" OR "confidence intervals" OR "Statistical Estimation" OR "hypothesis tests" OR "hypothesis testing" OR "significant coefficient" OR "significant parameter" OR "standard error" OR "Statistical Analysis" OR "intermediary statistics" OR "intermediate statistics") NOT (Abstract: "distributed lag" OR Abstract: "normally distributed" OR Abstract: sensor* OR Abstract: electric* OR Abstract: signal*)

## 2.2.5. Grey literature

As one of the exclusion criteria is to exclude unpublished studies, no research was conducted among grey literature.

## 2.3. Selection process

After removing duplicates, a manual review of the references selected by the databases will be done to select relevant references to answer the abovementioned research question. A two-stages selection process will be applied in this study.

To ensure consistency, all reviewers involved will meet before the first stage of selection, after the first stage of selection, midpoint through the second stage, and at the end of the second stage. Also, inclusion criteria were discussed among reviewers to ensure shared definitions.

### 2.3.1. Stages of the selection

Selection 1: Titles and Abstracts

All titles and abstracts of the references identified by the research strategy will be evaluated by one author (MPD). As this step is performed by one person only, only references unrelated to the research question will be automatically discarded, as well as the references that undoubtedly do not meet inclusion criteria.

Selection 2: Full text

The full texts of the references selected in selection 1 will be reviewed by two authors (MPD and SL). In the case of opposite opinions from the two initial reviewers, the reviewers will discuss and, if needed, a third author (FCL) will oversee a third review to decide.

Additional strategy

The list of references of all included articles after selection will be assessed to include articles that could not have been screened because of specific keywords.

### 2.3.2. Inclusion criteria

The following criteria were used to proceed to selection. Exclusion was assessed when at least one of the exclusion criteria was met or when at least one of the inclusion criteria was not met.

| *Criteria topic* | **Inclusion criteria** | **Exclusion criteria** |
|---|---|---|
| 1. Vertically partitioned data | This paper/study presents a solution to perform inferential statistics with vertically partitioned data. *Examples of papers that would not meet the criteria: the method is* | |

|  |  |  |  |
|---|---|---|---|
|  | *presented on horizontally partitioned data only; or the method requires pooling all line-level data.* |  |  |
| 2. Inferential Statistics |  |  | The paper/study does not specifically address inferential statistics (Confidence intervals, Hypothesis testing or Asymptotic normality result). *e.g., the focus is not on estimation and/or confidence intervals and/or hypothesis testing.* |
| 3. Methodological contribution |  |  | The paper/study does not provide a new methodological contribution. *e.g., the study is solely an application of a previously developed and presented method.* |
| 4. Published study |  |  | The paper/study has not been published. |
| 5. Language |  |  | The full-text is not available in English or French. |

## 2.4. Data-charting

A data-charting form was collectively developed, and extraction will be addressed manually. Data will be independently extracted by two authors (MPD and SL) for all studies. In the case of opposite opinions from the two initial reviewers, the reviewers will discuss and, if needed, a third author (FCL) will oversee a third review to decide. The authors will meet to update the data-charting form if needed.

The following information regarding the research questions were extracted in the process. Data-charting is seen as an iterative process, and changes will be done if needed to adjust and improve the extraction process. Covidence systematic review tool (www.covidence.org) was used.

*Figure 1. Extraction template from Covidence (1 of 2)*

# General information

# Characteristics of included studies

## Methods

Type of Model/Inference task

1. ☐ Linear Regression
2. ☐ Logistic Regression
3. ☐ Cox Regression
4. ☐ Other types of parametric/semi-parametric regression
5. ☐ Non-parametric regression
6. ☐ Tests non-specific to a regression setting
7. ☐ Other

Additional comments regarding the inference tasks conducted

Types of coordinating center (sometimes refered to as third party)

1. ○ None (all nodes have the same role)
2. ○ One of the nodes has a specific role
3. ○ External party to the nodes
4. ○ Both possibilities for a third party : External party or the nodes
5. ○ Not applicable / Unknown

Is there any mention of potential applications to health analytics ?

1. ○ Yes
2. ○ No
3. ○ Not applicable / Unknown

Number of communications ?

1. ○ Single (including back-and-forth)
2. ○ Iterative scheme
3. ○ Other / Unknown

Comments regarding the communication scheme (if applicable)

Partition setting

*Figure 2. Extraction template from Covidence (2 of 2)*

1. ○ Strictly Vertically Partitioned data
2. ○ Mixed Partitions (Vertical AND Horizontal)
3. ○ Unknown

Main Methodological Setting for Vertical Computations

1. ○ Mostly relies on Secure Matrix Computations
2. ○ Mostly relies on encryption (homomorphic or other)
3. ○ Mostly relies on Vertically Separable Quantities
4. ○ Other / Unknown

Additional comments regarding the methodological setting (e.g. uses a second strategy or a mix of strategies)

## PRIVACY

Does the method require sharing one or multiple raw variables (e.g. outcome variable) ?

1. ○ Yes
2. ○ No
3. ○ Not applicable / Unknown

Is Privacy/Confidentiality/Security mentioned ?

1. ○ Yes
2. ○ No

Do the authors claim any Privacy/Confidentiality/Security protection (even only for some settings) ?

1. ○ Yes, based on arguments OTHER than differential privacy.
2. ○ Yes, based on arguments regarding differential privacy.
3. ○ No.
4. ○ Not applicable / Unknown

Check all applicable statements

1. ☐ The paper discusses not sharing raw data.
2. ☐ The paper discusses reverse-engineering processes.
3. ☐ The paper provides an extensive privacy assessment, considering most reverse-engineering possibilities.
4. ☐ The paper provides privacy assessment based on encryption litterature.

Additional comments regarding the confidentiality / privacy ?

Other comments ?

*Multi-Site Health Research Integrating Complementary Data Sources: A Scoping Review of Statistical Inference Methods for Vertically Partitioned Data* - APPENDIX B. Supplementary tables

Table B.1. List of all 30 included articles[1]

| Ref[2] | Authors | Title | Year | Publication |
|---|---|---|---|---|
| 44 | Cock, M. de; Dowsley, R.; Nascimento, A. C. A.; Newman, S. C. | Fast, Privacy Preserving Linear Regression over Distributed Datasets based on Pre-Distributed Data | 2015 | Proceedings of the 8th ACM Workshop on Artificial Intelligence and Security |
| 60 | Movahedi, M.; Case, B. M.; Honaker, J.; Knox, A.; Li, L.; Li, Y. P.; Saravanan, S.; Sengupta, S.; Taubeneck, E. | Privacy-Preserving Randomized Controlled Trials: A Protocol for Industry Scale Deployment | 2021 | Proceedings of the 2021 on Cloud Computing Security Workshop |
| 50 | Hector, E. C.; Song, P.X.-K. | Doubly distributed supervised learning and inference with high-dimensional correlated outcomes | 2020 | Journal of Machine Learning Research |
| 51 | Hector, E.C.; Song, P.X.-K. | Joint integrative analysis of multiple data sources with correlated vector outcomes | 2022 | Annals of Applied Statistics |
| 49 | Dai, W.; Jiang, X.; Bonomi, L.; Li, Y.; Xiong, H.; Ohno-Machado, L. | VERTICOX: Vertically Distributed Cox Proportional Hazards Model Using the Alternating Direction Method of Multipliers | 2022 | IEEE Transactions on Knowledge and Data Engineering |
| 59 | Annamalai, M.S.M.S.; Jin, C.; Aung, K.M.M. | Communication-Efficient Secure Federated Statistical Tests from Multiparty Homomorphic Encryption | 2022 | Applied Sciences (Switzerland) |
| 25 | Imakura, A.; Tsunoda, R.; Kagawa, R.; Yamagata, K.; Sakurai, T. | DC-COX: Data collaboration Cox proportional hazards model for privacy-preserving survival analysis on multiple parties | 2023 | Journal of Biomedical Informatics |
| 48 | Kim, J.; Li, W.; Bath, T.; Jiang, X.; Ohno-Machado, L. | VERTIcal Grid lOgistic regression with Confidence Intervals (VERTIGO-CI) | 2021 | Proceedings – AMIA Joint Summits on Translational Science |
| 53 | Snoke, J.; Brick, T.R.; Slavković, A.; Hunte, M.D. | Providing accurate models across private partitioned data: Secure maximum likelihood estimation | 2018 | Annals of Applied Statistics |

| 58 | Ricci, S.; Domingo-Ferrer, J.; Sánchez, D. | Privacy-preserving cloud-based statistical analyses on sensitive categorical data | 2016 | International Conference on Modeling Decisions for Artificial Intelligence |
|---|---|---|---|---|
| 55 | Kikuchi, H.; Sato, T.; Sakuma, J. | Privacy-preserving hypothesis testing for the analysis of epidemiological medical data | 2014 | Proceedings - International Conference on Advanced Information Networking and Applications, AINA |
| 47 | Duverle, D.A.; Kawasaki, S.; Yamada, Y.; Sakuma, J.; Tsuda, K. | Privacy-preserving statistical analysis by exact logistic regression | 2015 | Proceedings - 2015 IEEE Security and Privacy Workshops |
| 35 | Liu, M.-C.; Zhang, N. | A cryptographic solution to privacy-preserving two-party sign test computation on vertically partitioned data | 2012 | Advanced Materials Research |
| 34 | Slavkovic, A.B.; Nardi, Y.; Tibbits, M.M. | Secure logistic regression of horizontally and vertically partitioned distributed databases | 2007 | Proceedings - IEEE International Conference on Data Mining, ICDM |
| 43 | Karr, A.F.; Lin, X.; Sanil, A.P.; Reiter, J.P. | Privacy-preserving analysis of vertically partitioned data using secure matrix products | 2009 | Journal of Official Statistics |
| 45 | Fienberg, S.E.; Nardi, Y.; Slavković, A.B. | Valid statistical analysis for logistic regression with multiple sources | 2009 | Annual Workshop on Information Privacy and National Security |
| 57 | Liu, M.-C.; Zhang, N. | A solution to privacy-preserving two-party sign test on vertically partitioned data (P22NSTv) using data disguising techniques | 2010 | ICNIT 2010 - 2010 International Conference on Networking and Information Technology |
| 42 | Karr, A.F.; Lin, X.; Sanil, A.P.; Reiter, J.P. | Secure statistical analysis of distributed databases | 2006 | Statistical Methods in Counterterrorism: Game Theory, Modeling, Syndromic Surveillance, and Biometric Authentication |
| 37 | Du, W.; Han, Y.S.; Chen, S. | Privacy-preserving multivariate statistical analysis: Linear regression and classification | 2004 | SIAM Proceedings Series |
| 36 | Du, W.; Atallah, M.J. | Privacy-preserving cooperative statistical analysis | 2001 | Proceedings - Annual Computer Security Applications Conference, ACSAC |
| 14 | Kamphorst, B.; Rooijakkers, T.; Veugen, T.; Cellamare, M.; Knoors, D. | Accurate training of the Cox proportional hazards model on vertically-partitioned data while preserving privacy. | 2022 | BMC medical informatics and decision making |
| **41** | Kikuchi, H.; Hamanaga, C.; Yasunaga, H.; Matsui, H.; Hashimoto, H. | Privacy-Preserving Multiple Linear Regression of Vertically Partitioned Real Medical Datasets | 2017 | IEEE 31st International Conference on Advanced Information Networking and Applications (AINA) |
| **40** | Kikuchi, H.; Hashimoto, H.; Yasunaga, H.; Saito, T. | Scalability of Privacy-Preserving Linear Regression in Epidemiological Studies | 2015 | IEEE 29th International Conference on Advanced Information Networking and Applications |
| **46** | Kikuchi, H.; Yasunaga, H.; Matsui, H.; Fan, C. -I. | Efficient Privacy-Preserving Logistic Regression with Iteratively Re-weighted Least Squares | 2016 | 11th Asia Joint Conference on Information Security (AsiaJCIS) |
| **56** | Wu, F.; Xi, B. | Differentially Private Causal Inference Under Hierarchical Design | 2023 | IEEE International Conference on Data Mining Workshops (ICDMW) |
| **39** | Hall, R.; Fienberg, S.E; Nardi, Y. | Secure Multiple Linear Regression Based on Homomorphic Encryption | 2011 | Journal of official statistics |
| **33** | Fienberg, S. E; Karr, A. F; Nardi, Y.; Slavkovic, A.B. | Secure Logistic Regression with Multi-Party Distributed Databases | 2007 | Proceedings of the 56th Session of the ISI |
| **38** | Reiter, J. P.; Kohnen, C.N.; Karr, A. F; Lin, X.; Sanil, A. P | Secure Regression for Vertically Partitioned, Partially Overlapping Data | 2004 | Proceedings of the American Statistical Association |
| **52** | Hector, E. C.; Song, P. X.-K. | A Distributed and Integrated Method of Moments for High-Dimensional Correlated Data Analysis | 2021 | Journal of the American Statistical Association |
| **54** | Snoke, J.; Brick, T.; Slavković, A. | Accurate Estimation of Structural Equation Models with Remote Partitioned Data | 2016 | Privacy in Statistical Databases |

[1]Search queries were run on July 18$^{th}$, 2026.
[2]Reference number as it appears in main text.

Table B.2. Primary methodological technique among included articles

| Primary methodological technique for the vertical setting | Number of articles[1] | Article |
|---|---|---|
| • Vertically separable quantities | 4 | [1–4] |
| • SMC | 11 | [5–15] |
| • Encryption | 13 | [14,16–27] |
| • Other | 3 | [28–30] |

[1]Total number does not necessarily equal 30, because an article might present methods use more than one primary methodological technique.

**Additional note:** We observed three techniques for the vertical setting: encryption-based methods, methods based on vertically separable quantities and methods based on secure multiparty computations (SMC). Encryption protocols rely on key-generation; quantities are encrypted using the key before being shared to allow operations, after which results are sent back and decoded using the key. In the case of vertically separable quantities, the quantity of interest naturally separates piecewise among data nodes; every node computes the local version of this quantity, which, once aggregated through a sum or another operation, exactly recreates the aimed global quantity. For example, the gram matrix of a covariate matrix $X$, in which columns represent variables, defined as $XX^T$, is vertically separable, because it is the sum of all local gram matrices computed from local covariates matrices (see [2]). Finally, SMC protocols draw partly on the two previous methodologies; randomness is incorporated in the quantity over which operations need to be run in a prescribed way such that, once the collaboration computations are completed, the randomness is canceled out (e.g. multiplying by a matrix in a null-space). One article included both SMC and encryption procedures without emphasizing on one or another [14]. The three articles classified as *Other* [28–30] could have been classified as separable quantities but pertained to a specific setting where the vertical component arises from multi-dimensional outcomes, which could also be interpreted as horizontally separable and were therefore treated as exceptions.

**References (Appendix B)**